\documentclass[12pt,preprint]{aastex}
\begin{document}
\title{The Extreme Ultraviolet Deficit and Magnetically Arrested Accretion in Radio Loud Quasars}
\author{Brian Punsly\altaffilmark{1}}
\altaffiltext{1}{1415 Granvia Altamira, Palos Verdes Estates CA, USA
90274 and ICRANet, Piazza della Repubblica 10 Pescara 65100, Italy,
brian.punsly1@verizon.net or brian.punsly@comdev-usa.com}
\begin{abstract}
The Hubble Space Telescope composite quasar spectra presented in
Telfer et al. show a significant deficit of emission in the extreme
ultraviolet (EUV) for the radio loud component of the quasar
population (RLQs), compared to the radio quiet component of the
quasar population (RQQs). The composite quasar continuum emission
between 1100 \AA\, and $\sim$580 \AA\, is generally considered to be
associated with the innermost regions of the accretion flow onto the
central black hole. The deficit between 1100 \AA\, and 580 \AA\, in
RLQs has a straightforward interpretation as a missing or a
suppressed innermost region of local energy dissipation in the
accretion flow. It is proposed that this can be the result of
islands of large scale magnetic flux in RLQs that are located close
to the central black hole that remove energy from the accretion flow
as Poynting flux (sometimes called magnetically arrested accretion).
These magnetic islands are natural sites for launching relativistic
jets. Based on the Telfer et al. data and the numerical simulations
of accretion flows in Penna et al., the magnetic islands are
concentrated between the event horizon and an outer boundary of
$<2.8 M$ (in geometrized units) for rapidly rotating black holes and
$< 5.5M$ for modestly rotating black holes.
\end{abstract}


\section{Introduction}
The nature of the causative agent that makes some quasars radio loud
(RLQs) has challenged astrophysicists for more than 50 years. It
became clear early on that the optical/ultraviolet (UV) spectra of
RLQs and radio quiet quasars (RQs) are very similar
\citep{ste91}\footnote{There are notable small subclasses of objects
that are distinct. Some RLQs have relativistic jets that propagate
close to the line of sight (blazars) and the Doppler enhanced power
law continuum can be significant. There are also rare objects with
very broad ultraviolet absorption lines, these are almost
exclusively RQs. However, in both classes these effects obfuscate a
background thermal component that is very similar to other
quasars.}. Attempts to look for subtle differences involved
statistical studies of optical and UV emission line strengths and
widths \citep{bor92,bor02,cor94,cor96,bro94}. These emission regions
are far from the central engine, $\sim$ $10^{3}$ - $10^{4}$ larger
than the central black hole radius, so it is not clear what they
tell us as a second order indicator of conditions in the jet
launching region \citep{gue13}. Are they related to the fueling
mechanism for radio loudness, the ionization continuum or jet
propagation? Consequently, this research path has provided very
little understanding of the jet launching mechanism. Seemingly more
relevant to the physics of jet launching, the extreme ultraviolet
(EUV) continuum, $\lambda < 1100$ \AA\,, is created orders of
magnitude closer to the central engine and RLQs display a
significant EUV continuum deficit relative to RQs \citep{tel02},
\textbf{T02} hereafter.
\par In the following, evidence is presented that connects the
EUV deficit to magnetically arrested accretion (MAA) in the
innermost accretion flow of RLQs. The motivation for exploring this
interpretation is that it not only explains the second order effect
of an EUV deficit, but it also provides a mechanism for the first
order difference between RLQs and RQs; namely arresting the flow
with large scale magnetic flux is a natural way to launch the jets
responsible for the radio emission. This argument is laid out as
follows. Section 2 considers the EUV emission in the context in the
standard interpretation of a quasar as emission from an optically
thick thermal gas that accretes onto a black hole. Section 3 reviews
the notion of MAA. Based on the assumption that the EUV deficit is a
consequence of thermal gas being displaced by islands of large scale
magnetic flux, the distribution of said islands is determined from
both numerical and theoretical models of accretion flows.
\section{The Thermal Interpretation of Quasar Spectra}
It was convincingly demonstrated in \citet{lyn71,sha73,nov73} that the
intense blue/UV light associated with the quasar phenomenon was
likely the optically thick thermal emission from viscous dissipation
of accreting gas onto a supermassive black hole. The connection
between these accretion models and observation was made in
\citet{mal83,szu96} where quasar spectra in the the optical to far
UV were approximated by accretion disks spectra.
\begin{figure}
\includegraphics[width=170 mm, angle= 0]{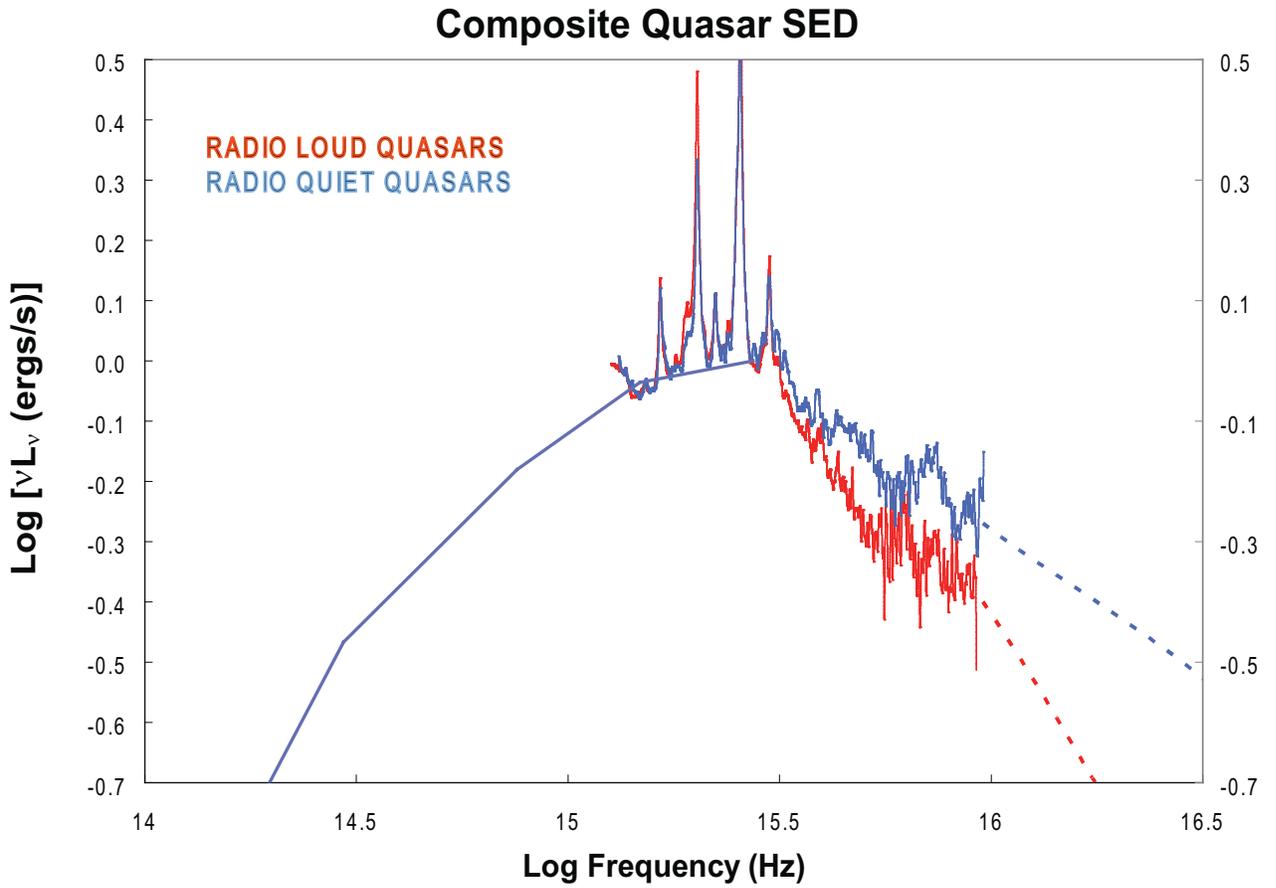}
\caption{The quasar composite continuum accretion disk SED. The
black lines represent the common UV/optical/IR continuum. The blue
and red plots are the EUV composite spectra for RQs and RLQs from
\textbf{T02}, respectively.}
\end{figure}
Our understanding of accretion disks around black holes is far from
complete. Thus, this effort strives to reach conclusions that do not
depend on a particular accretion model. Determination of the EUV
spectrum beyond the far UV turnover had to await space-based
observations of intermediate redshift quasars since ground-based
observations of high redshift objects, in which the EUV is
redshifted into the optical band, are unsuitable due to
contamination by the Ly$\alpha$ forrest \citep{zhe97}.
\par The EUV deficit in RLQs was originally found in \citet{zhe97} and confirmed with a much larger
sample of spectra in \textbf{T02}. Figure 1 shows the composite RLQ
(red) and RQ (blue) EUV spectra, 1100 \AA\, to 300 \AA\,, from
Hubble Space Telescope (HST) observations of 332 spectra of 184
objects with redshift, $z>0.33$ \textbf{T02}\footnote{The newer HST
spectra in \citet{ste14} have a span $<1/2$ of the G160L spectra
(that was commonly used in \textbf{T02} to bridge the far UV and
EUV) and are not wide enough to use in Figure 2. The small number
(8) of narrow span RLQ spectra entirely within the EUV, above the
noisy spectral break at $\approx 550 \AA$ of Figure 1 ($1100\AA\ <
\lambda < 550 \AA $), render statistical comparison to RQs
insignificant. The FUSE composite in \citet{sco04} has only two RLQs
with coverage below 700 \AA\,.}. The quasar accretion disk composite
in Figure 1 is based on the \citet{lao97} composite, but it is
updated with the optical and IR composite quasar estimates in
\citet{dav11}. They note the important point that the IR (dust)
local maximum should be neglected in estimates of the accretion flow
SED because it is likely to be the result of accretion disk emission
that is reprocessed on larger scales. The UV composite is updated
based on the data and discussion in \textbf{T02}. The small
difference in the UV continuum between the RLQ and RQ composites in
\citet{lao97} does not exist (see Figure 10 of \textbf{T02}) in the
larger HST sample. The common continuum at frequencies below the far
UV for the accretion disk composite of RQs and RLQs is represented
by a black piecewise power-law fit. The EUV is the data of interest,
so it is plotted explicitly as opposed to the piecewise power-law
estimates elsewhere. The EUV data is normalized to 0 at 1100 \AA\,
as in \textbf{T02}. The SEDs above $6\times 10^{15}$ Hz show a
spectral break, but they are very noisy and are not considered
reliable in this region. At frequencies above $10^{16}$ Hz, where
the \textbf{T02} data ends, the SED is very uncertain and is
indicated by faint dotted lines. These extend to the soft X-ray
values (relative to the peak of the SED) from \citet{lao97}. Since
the X-ray luminosity is not considered optically thick thermal
emission from the accretion flow and its value relative to the EUV
is quite uncertain, it will not be plotted or considered in detail
in the following.
\par It should be noted that both numerical and analytic models of
optically thick accretion flows contain the following elements: the
effective temperature increases as the radius decreases and the
luminosity from each annular ring reaches a maxima near the black
hole and decreases inward of this, i.e., the luminosity of the
accreting gas fades before being swallowed by the black hole
\citep{zhu12}. Thus, we expect that the maximum of the SED is not
representative of the maximum temperature of the accretion flow, but
there are higher temperature contributions to the SED beyond the far
UV turnover from optically thick thermal gas. Thus, the rapidly falling SED in
the EUV band is the electromagnetic signal of the innermost
optically thick region. Consider this in the context of the
broadband composite of Figure 1. The continuum of the thermal
component (frequencies below $6\times 10^{15}$ Hz) of RLQs and RQs
are indistinguishable except for emission from the innermost
accreting optically thick gas. Thus, the difference in the EUV
emission between RLQs and RQs likely arises from suppressed emission
in the innermost region of the accretion flow in RLQs of what is
otherwise a similar accretion flow to that found in RQs.
\begin{figure}
\includegraphics[width=170 mm, angle= 0]{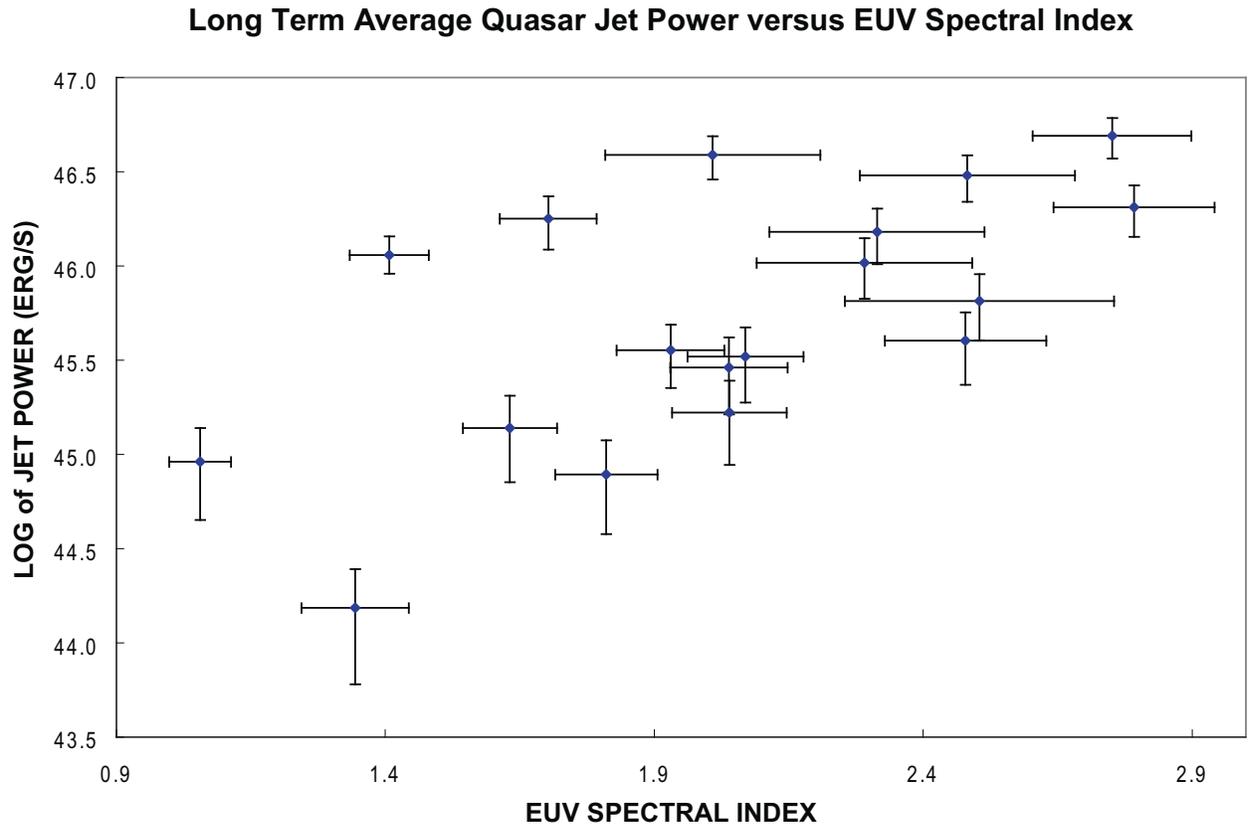}
\caption{A scatter plot of the estimated jet power and the EUV
spectral index in frequency space (a value of 1 is a flat SED).}
\end{figure}
\section{The EUV Deficit and Magnetically Arrested Accretion}
The viscous dissipation that heats the plasma in accretion flows
(and therefore the source of modified black body radiation) is
produced as a consequence of the magneto-rotational instability
(MRI) in 3-D numerical simulations \citep{pen10,dev03}. The
discussion of the last section begs the question, what physical
process in the innermost accretion flow can suppress MRI in RLQs?
The answer might be MAA, \citet{igu03, igu08}, or its variants MCAFs
(magnetically choked accretion flows), \citet{mck12} and MADs
(magnetically arrested disks), \citet{tch11,tch12}. The accretion
flow in these simulations is perforated by large scale magnetic flux
tubes, magnetic islands. The islands of magnetic flux that arrest
the accretion flow also suppress the MRI induced dissipation in
these regions. The magnetic flux tubes torque the plasma and enhance
the overall mass accretion rate. The angular momentum is converted
to electromagnetic form and removed vertically from the accretion
flow as a jet \citep{igu08}. Without loss of generality, consider an
annular ring in the \citet{sha73} accretion disk, $r_{1}<r<r_{2}$.
Angular momentum is removed by viscous stress in the fluid element
at a rate
\begin{equation}
T_{\phi\; ;\nu}^{\,\nu} =0 \,\Longrightarrow \, \dot{L}
=\dot{m}(\Omega(r_{2}) r_{2}^{2} - \Omega(r_{1}) r_{1}^{2}) =
\int{r^{-2}(r^{2}T_{\phi}^{r\;\,\mathrm{visc}})_{,r} dV}\;,
\end{equation}
where $T_{r \phi}^{\mathrm{visc}}$ is the viscous stress, $\dot{m}$
is the accretion rate and $\Omega (r)$ is the angular velocity. Now
consider the existence of magnetic islands that fill a fraction,
$f_{V}$, of the volume of the ring, $V$, and penetrate a fraction,
$f$, of the top and bottom surface areas of the annular volume,
$SA$.  The volume of magnetic islands is $V_{MI}$ and its complement
in $V$ is $V_{MI}^{C}$, $f_{V} = \int{dV_{MI}}/V$. The surface area
elements of the top and bottom faces are, $dS\!A_{MI}$ and
$dS\!A_{MI}^{C}$, respectively. The angular momentum equation
becomes
\begin{equation}\dot{L} =\dot{m}(\Omega(r_{2}) r_{2}^{2} - \Omega(r_{1}) r_{1}^{2})  = \int{r^{-2}(r^{2}T_{\phi}^{r\;\,\mathrm{visc}})_{,r} dV_{MI}^{C}} + \int{(-rB^{\phi}B^{z}/(8\pi))_{,z} dV_{MI}} ,
\end{equation}
 where, $B^{\phi}$ and $B^{z}$ are the azimuthal and vertical magnetic field components, respectively.
For $\dot{m}$ fixed in Equations (2) and (3), even though the volume
of plasma experiencing viscous dissipation is reduced, accretion
proceeds at an equal rate in the magnetically arrested state.

\par Simulated MAD accretion flows are subsonic and therefore do not produce significant gas heating from
shocks \citep{mck12,pun14}. Thus, the reconnection of the locally
tangled field driven by the MRI is the primary source of heat
creation at the boundary of the magnetic islands and in the
accreting gas. However, the interior of the magnetic islands are not
regions of local MRI driven heating. The total volume available for
MRI induced viscous heating is reduced by the magnetic islands.
Therefore, the MRI suppression in magnetically arrested flows
indicate states of lower radiative efficiency relative to standard
accretion states without magnetic islands. The magnetic islands
radiate Poynting flux along the magnetic field lines. Again consider
the annular ring above. The total energy flux, $Q =
Q^{\mathrm{visc}} + Q^{\mathrm{jet}}$, has two components, where
$Q^{\mathrm{visc}}$ is the flux of radiation as in a standard
accretion disk and $Q^{\mathrm{jet}}$ is primarily poloidal Poynting
flux along the magnetic field direction, $S^{P}$ \citep{igu08}.
Similarly, the total luminosity $P$ of the ring also has two
components:

\begin{equation}P = \int{Q\,dA} = (1/2)\int{r (d\Omega/dr)T_{r \phi}^{\mathrm{visc}}dz\,dS\!A_{MI}^{C}} + \int{S^{P} dS\!A_{MI}} \;.
\end{equation}
The first term on the RHS of Equation (3) is the usual term from
standard accretion theory that gives rise to the radiation (such as
EUV). The second term is the vertical jet emission. For any
approximately axisymmetric MHD Poynting flux dominated jet,
regardless of the source, the total integrated electromagnetic
poloidal energy flux is
\begin{equation}
\int S^{P} \mathrm{d}A_{_{\perp}}
=(\Omega_{F}/c)\int{(-rB^{\phi}B^{z}/(8\pi))_{,z} dV_{MI}} \approx
k\frac{\Omega_{F}^{2}\Phi^{2}}{2\pi^{2} c}\;,
\end{equation}
where $\Phi$ is the total magnetic flux enclosed within the jet,
$\mathrm{d}A_{_{\perp}}$ is the cross-sectional area element and $k$
is a geometrical factor that equals 1 for a uniform highly
collimated jet \citep{pun08}. Thus, not only do the magnetic islands
of large scale poloidal flux in the inner accretion flow suppress
radiation from this region, but they provide a source of Poynting
flux (power for the jet) as they orbit around the black hole with an
angular velocity, $\Omega_{\mathrm{F}}$. The local physics that
produces the turbulent viscosity, $\eta_{t}$, in $V_{MI}^{C}$ is
unchanged from standard accretion and therefore so is $T_{r
\phi}^{\mathrm{visc}}= \eta_{t}r(d\Omega/dr)$. Thus, from Equations
(1) - (4), in the magnetically arrested case, the radiative
luminosity is $\approx 1-f$ of what it would be for standard
accretion with the same mass accretion rate.
\par If MAA is the source of the EUV
deficit, one would expect a correlation (perhaps weak) between the
EUV spectral index, $\alpha_{\mathrm{EUV}}$, and jet power within
the RLQ population. In Figure 2, $\alpha_{\mathrm{EUV}}$
($F_{\nu}\sim\nu^{-\alpha_{\mathrm{EUV}}}$) derived from individual
spectra in the HST archives (downloaded through MAST) is plotted
against estimates of the long term time average of the jet power,
$\overline{Q}$. In order to get a meaningful estimate of
$\alpha_{\mathrm{EUV}}$, a range of at least 700 \AA\, to 1100 \AA\,
in the quasar rest frame was needed to extract the continuum from
the numerous broad emission lines documented in \textbf{T02}.
Therefore, a redshift of $z>0.63$ is required. Troughs from Lyman
limit systems (LLS) were removed by assuming a single cloud with a
$\nu^{-3}$ opacity. This was considered acceptable if the power law
above the LLS could be continued smoothly through the corrected
region. If there were many strong absorption systems or an LLS that
compromised a broad emission line, this simple procedure was deemed
inadequate for continuum extraction with the available data and the
spectrum was eliminated from the sample. A small correction for the
Lyman valley was also made per the methods of \citet{zhe97}.
Additionally, if there was evidence of a blazar synchrotron
component contribution to the continuum (high optical polarization
or variability, superluminal motion or gamma ray activity), the
underlying accretion disk continuum was considered too uncertain for
the sample. \par The most reliable methods of estimating
$\overline{Q}$ are based on the the optically thin emission from
relaxed radio lobes \citep{wil99}. Thus, all sources in the sample
needed proof of extended emission on scales larger than the host
galaxy so that the lobes can relax ($> 20$ kpc). The proof was
derived from archival high resolution interferometry images made
between 0.408 GHz and 5 GHz. The HST and radio selection criteria
resulted in a total of 18 sources for the sample. The optically thin
emission was estimated based on 151 MHz - 178  MHz flux densities
(if available) and the lobe fluxes from the radio images. The
largest spread in the estimates of $\overline{Q}$, based on
optically thin extended emission, are bounded on the high side by
the \citet{wil99} estimate for their parameter, \textbf{f} = 20, and
on the low side by \citet{pun05}, which assumes that the lobes are
inertially dominated. These two extremes are used to generate the
errors bars on $\overline{Q}$ in Figure 2 \citep{pun14}. The
correlation in the scatter in Figure 2 is statistically significant
at the 0.987 level by a Spearman rank correlation test, while the
correlations of $\alpha_{\mathrm{EUV}}$ with z and spectral
luminosity, $\lambda L_{\lambda}(1100 \AA)$, are significant at the
0.842 and 0.720 levels, respectively for the same sample. The large
scatter in Figure 2 is expected on many grounds: the ejections
producing $\overline{Q}$ are not contemporaneous with the the HST
spectrum, as well as variations in the quasar host EUV absorption,
$\dot{m}$, black hole masses, ($M$), and spins ($a/M$).
\par Since a magnetically arrested innermost accretion flow
naturally explains the suppressed EUV and jet production, it is of
interest to estimate the size of the region of suppressed radiation.
From, the composites in Figure 1, the EUV deficit in RLQs is
$\approx 0.045L_{\mathrm{bol}}$ of the optically thick accretion
flow\footnote{The emission line contribution to $L_{\mathrm{bol}}$
is chosen to be 25\% of the optical/UV luminosity \citep{zhe97}. The
X-ray contribution to $L_{\mathrm{bol}}$ is ignored as discussed in
Section 2. Ignoring the X-ray luminosity of the accretion flow
proper will affect the estimates by $<$ 10\% \citep{lao97,dav11}.}.
Consider excising a fraction, $f$, of the innermost accretion flow
in various models of accretion disks in Figure 3. The simulations of
accretion disks in \citet{pen10} include luminosity from the plunge region inside of the innermost stable orbit (ISCO) and are
parameterized by $a/M$ and the disk thickness defined by their
parameter, $h/r$. The \citet{nov73}, NT, models do not include a
plunge region. Note that the putative magnetically arrested region
must be concentrated at the smallest radii, since the EUV is
suppressed in RLQ composite of Figure 1, but not the UV and optical.
The plausible range of, $0.3 < f < 0.9$, near the black hole is
based on the simulations presented in \citet{igu08,pun09}. The
putative magnetic islands would have to be concentrated between the
event horizon and an outer boundary of $<2.8 M$ if $a/M=0.98$ and
$<5.5M$ if $a/M=0.7$ to account for the 4.5\% luminosity suppression
in RLQs.
\begin{figure}
\includegraphics[width=170 mm, angle= 0]{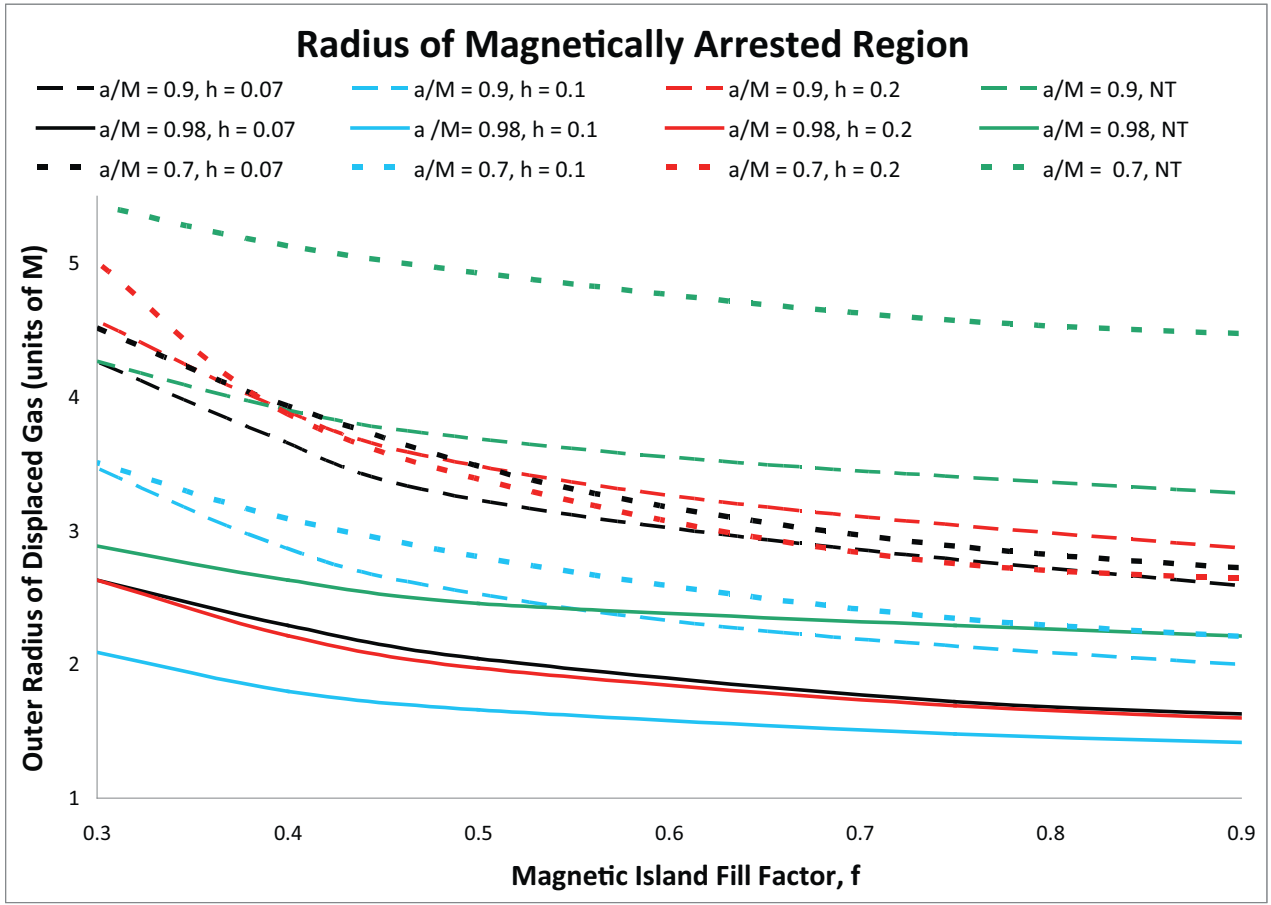}
\caption{The size of the magnetically arrested region required for
the EUV deficit in RLQs versus the filling factor for magnetic flux,
f, for various models from \citet{pen10}.}
\end{figure}
\par It should be noted that, unlike the simulations in
\citet{igu08,pun09}, the simulations in \citet{mck12,tch11,tch12}
that are heavily seeded with large scale magnetic flux are devoid of
magnetic islands this close to the event horizon. This is evidenced
by the claim in \citet{mck12} that no significant Poynting flux
emerges from this region (see Equation 4, above) as well as the
linked online videos of the simulations. The videos show the
innermost significant magnetic island concentrations are around
$\sim 20M$ and they are extremely transient. This either means that
the interpretation of the EUV deficit presented here is wrong or the
simulations do not represent the magnetic flux evolution accurately.
Using fusion and solar physics as a guide, the latter seems quite
likely since these simulations are based on simple single fluid
ideal MHD \citep{yam07,mal09,bau13,thr12}. Even more important, the
two most relevant dynamic elements for large scale, poloidal,
magnetic flux evolution near the black hole, reconnection and
diffusion of mass onto the field lines, occur as a consequence of
numerical diffusion in the simulations and not an actual physical
process.
\section{Discussion}
In this paper, the EUV deficit in RLQ SEDs was argued to arise from
a deficit of optically thick thermal gas in the innermost accretion
flow. It was posited that islands of large scale magnetic flux near
the black hole, like those that occur in some numerical simulations,
would explain this missing volume of optically thick thermal gas and
also explain the radio jet launching mechanism. As a further
consequence it was argued in \citet{pun99} that the presence of
magnetic flux in the inner accretion disk will diminish the power of
radiation driven winds that appear to be common in RQs. It is not
claimed that this is the only explanation of the EUV deficit.
However, none of the alternatives naturally produce a radio jet.
Other explanations based on numerical and theoretical models
include, lower $a/M$ in RLQs (larger ISCO), a stronger quenching
wind in RLQs per the model of \citet{lao14} or higher $M$ and lower
$\dot{m}$ in RLQs. Using a sample of $>6000$ QSOs
\citet{mcl04} found that the mean $M$ of RLQs is 1.45 that of of RQs
with large scatter. In disk models, larger M shifts the peak of the
SED to lower frequency, therefore causing a decrease in the EUV.
However, it was shown in \citet{dav11,lao14}, based on PG quasars,
that black hole mass variations produce much smaller changes in the
far UV turnover region of the spectrum than expected from accretion
disk models. Thus the issue needs to be addressed empirically. In
Figure 4, the \textbf{T02} composites are overlayed. The location of
the SED peak and the curvature of the continuum long-ward of the
peak are indistinguishable contrary to the notion that a disk
temperature shift results in the EUV deficit. Furthermore, if the mass difference
is the source of the EUV deficit, the correlation in Figure 2 would
be a coincidence.
\begin{figure}
\includegraphics[width=170 mm, angle= 0]{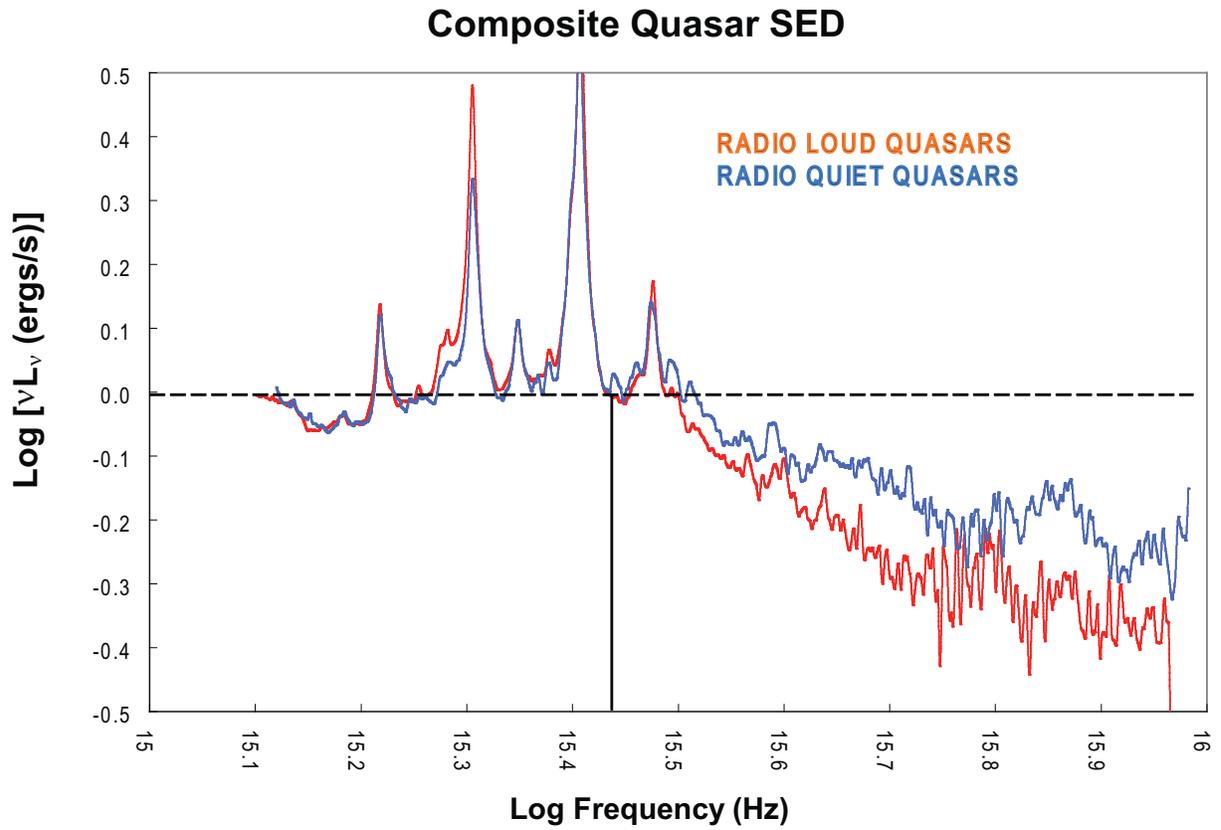}
\caption{The blue and red plots are the composite spectra for RQs
and RLQs from \textbf{T02}, respectively. The spectral peak and the
curvature of the continuum long-ward of the peak are virtually
identical. The composites are normalized to 0 at 1100\AA. That is
indicated by the black lines.}
\end{figure}

\par The basic idea presented here does not depend on any particular
accretion model. The observational data indicates that the only
significant difference in the optically thick thermal continuum
between the RQ and RLQ composite spectra is a deficit of the highest temperature gas
(the EUV) in RLQs. This must be created by the innermost optically
thick gas independent of the model. A plausible explanation is the
displacement of this gas by the large scale magnetic flux of the
radio jet at its launch site.
\begin{acknowledgements}
This work benefitted greatly from the input of a very knowledgable
referee who directed the effort towards many important topics that
were initially overlooked. I am also indebted to Michael Brotherton
who computed the RLQ and RQQ black hole masses from SDSS DR7, for
$0.9 < z <1.1$. Although not directly used due to space constraints,
this provided valuable insight. I am also thankful to Robert
Antonucci for helping me correct for the Lyman limit systems in the
HST spectra. I am extremely grateful to Matt Malkan who reviewed the
logic of the arguments presented and implied by the paper with me
and also had great insight into the proper interpretation of the HST
data.
\end{acknowledgements}

\end{document}